\documentclass[aps,pra,reprint,groupedaddress,amsmath,amssymb,preprintnumbers,showpacs,longbibliography]{revtex4-2}
\usepackage{graphicx}
\usepackage{bm}
\usepackage{dcolumn}
\usepackage{physics}
\usepackage{siunitx}
\newcommand{\be}{\begin{equation}}
\newcommand{\ee}{\end{equation}}
\newcommand{\ba}{\begin{eqnarray}}
\newcommand{\ea}{\end{eqnarray}}
\newcommand{\bs}{\begin{subequations}}
\newcommand{\es}{\end{subequations}}
\newcommand{\bw}{\begin{widetext}}
\newcommand{\ew}{\end{widetext}}

\newcommand{\knd}{k_{n,\text{D}}}

\begin{document}

\title{Nondipole photoelectron momentum shifts in strong-field ionization with mid-infrared laser pulses of long duration}

\author{Mads Middelhede Lund}
\affiliation{Department of Physics and Astronomy, Aarhus
University, DK-8000 Aarhus C, Denmark}

\author{Lars Bojer Madsen}
\affiliation{Department of Physics and Astronomy, Aarhus
University, DK-8000 Aarhus C, Denmark}

\date{\today}

\begin{abstract}
We consider atomic strong-field ionization in  the long-pulse limit for linearly polarized infrared laser fields.  We show how nondipole effects in the plane formed by the propagation and polarization directions lead to (i) a shift of the origin of the rings describing the energetically allowed final electron momenta for individual above-threshold ionization channels and (ii) a redistribution in the continuum population on each ring.
\end{abstract}

\maketitle

\section{\label{sec:introduction} Introduction}
Nondipole effects show up in strong-field ionization at near- and mid-infrared wavelengths and moderate to high intensities due to radiation pressure and magnetic field effects~\cite{Reiss1990,Reiss2008,Reiss2013,Reiss2014}. When the dipole approximation is accurate, the photoelectron momentum distribution (PMD) is symmetric under reflection in the laser polarization plane, i.e., under $k_x \rightarrow -k_x$ where $k_x$ is the momentum along the propagation direction $x$ (atomic units (a.u.) are used throughout). Nondipole effects break this symmetry.  

During the last decade, observations of nondipole-induced shifts of the PMD along the laser propagation direction in strong-field ionization experiments~\cite{Smeenk2011,Keller2014,Maurer2018,Haram2019,Doerner2019}  have fuelled interest as recently reviewed~\cite{Wang2020,Haram2020,Maurer2021}.  The experimental works~\cite{Smeenk2011,Keller2014,Haram2019,Maurer2018,Doerner2019}  used pulsed lasers of femtosecond duration. The finite pulse durations impeded resolving the individual above-threshold ionization (ATI) channels. The focus of the experiments~\cite{Smeenk2011,Keller2014,Maurer2018,Haram2019,Doerner2019} was therefore on the overall shift of the momentum distribution along the laser propagation direction. The shift can be parameterized in terms of well-known quantities such as ionization, $I_p$,  and ponderomotive, $U_p$, potentials~\cite{He2017}.   A saddle-point analysis in the short-pulse limit based on the strong-field approximation gives a shift of $I_p/(3c)$ ~\cite{Klaiber2013,Yakaboylu2013,Chelkowski2014,Chelkowski2015,He2017,Doerner2019,JensenPRA2020} and a consideration of the radiation pressure gives a shift of $U_p/c$~\cite{Reiss2013}. Note that  the Coulomb potential can shift momenta in the direction opposite to  that induced by the laser~\cite{Madsen2006,Bauer2017,Haram2019,Fritzsche2019,Doerner2019}.

 It is the purpose of this work to provide predictions for the shifts in the long-pulse limit, where the ATI channels can be resolved.  We find it advantageous to consider this regime because one can study the nondipole effect in the individual ATI channels, a point which seems to have escaped attention in the recent studies. We report nondipole-induced shifts of the center of the energy-conserving rings for a given ATI channel and illustrate redistribution of the continuum population on such a ring due to nondipole terms.  

The paper is organized as follows. In section II, the theory and analytical results are described. In section III,  results are given focusing on the mid-infrared regime. Section IV concludes. Atomic units are used throughout.

\section{\label{sec:theory} Theory}
In this section, we recall the nondipole strong-field-approximation Hamiltonian and the associated Volkov waves~\cite{JensenPRA2020}. We give expressions for the PMD in the nondipole and dipole approximations. Finally, energy constraints on the final momenta are discussed.

\subsection{Nondipole strong-field-approximation Hamiltonian}
Our approach was discussed in detail elsewhere~\cite{JensenPRA2020}, and recently applied to the related process of laser-assisted electron scattering~\cite{JensenJPB2020}, so the presentation here is brief. For linearly polarized light of low frequency and high intensity, it is accurate to consider the  approximate nondipole strong-field-approximation Hamiltonian for an electron in an electromagnetic field~\cite{JensenPRA2020}. In this approach, the leading-order nondipole magnetic field effects are included along the dipole-induced motion. The nondipole strong-field-approximation Hamiltonian reads in the velocity gauge (VG)~\cite{JensenPRA2020}
\begin{equation}
H_\text{ND,VG}^{\text{SFA}}=  H_\text{ND,VG}^{\text{SFA,L-e}}+V(\bm r).
\label{NDSFAVG}
\end{equation}
Here $V(\bm r)$ denotes the single-active-electron potential and 
\begin{equation}
H_\text{ND,VG}^{\text{SFA,L-e}}=\frac{(\bm p + \bm A'(t))^2 }{2} 
\label{eq:Hfdipol}
\end{equation}
describes the laser-electron (superscript L-e) interaction, with 
\begin{equation}
\bm A'( t) = A^{(0)}(t) \hat{\bm z} + \frac{A^{(0)}(t)^2}{2c} \hat{\bm x} \label{Aprime}
\end{equation}
the vector potential with linear polarization along $\hat{\bm z}$ and propagation along $\hat{\bm x}$. We model $\bm A^{(0)}(t)$ by the expression 
$\bm A^{(0)}(t) = \hat{\bm z} A^{(0)}(t) = \hat{\bm z} A_0 \sin(\omega t).$
The subscripts ND and VG in equations \eqref{NDSFAVG} and \eqref{eq:Hfdipol} refer to the inclusion of nondipole terms and the usage of the velocity gauge, respectively. The superscript SFA refers to the strong-field approximation. The superscript (0) on the vector potential reminds us that the vector potential only depends on time. We note that the Hamiltonian in equation \eqref{eq:Hfdipol} has been considered in detail in the high-intensity, high-frequency regime~\cite{shakeshaft,kylstra,Forre2014,Forre2015,Forre2016a,Forre2016b,Moe2018,Forre2020}.
 In the dipole approximation the term proportional to $1/c$ in equation \eqref{Aprime} is absent and equation \eqref{eq:Hfdipol} reduces to the dipole Hamiltonian.

\subsection{Nondipole strong-field-approximation Volkov wave function}
One verifies by substitution that 
\be
\psi^{V,\bm k}_\text{ND,VG}(\bm{r},t) = \frac{1}{(2\pi)^{3/2}} e^{i\bm k \cdot \bm r -i \int^t dt' (\bm k + \bm A'(t'))^2/2}
\label{eq:VolkovPG1}
\ee
is a solution to the time-dependent Schr\"{o}dinger equation, $i \partial_t \psi^{V,\bm k}_\text{ND,VG}(\bm{r},t) =   H_\text{ND,VG}^{\text{SFA,L-e}}(t) \psi^{V,\bm k}_\text{ND,VG}(\bm{r},t)$, with the Hamiltonian from equation \eqref{eq:Hfdipol}. The wave function in equation \eqref{eq:VolkovPG1} is the VG nondipole strong-field-approximation Volkov wave function~\cite{JensenPRA2020}.  In the long-pulse limit and to leading-order in $1/c$, it may be expressed as 
\begin{align}
\psi^{V,\bm k}_\text{ND,VG}(\bm{r},t) &= \frac{1}{(2 \pi)^{3/2}} \exp\left( i \bm{k} \cdot \bm{r} \right) \nonumber \\ \times \sum_{n=-\infty}^\infty  & e^{-i(k^2/2 + U_p' + n\omega)t} (- i)^n J_n \left( \alpha_0  k_z, \frac{U_p'}{2 \omega} \right). \label{NDVGV}
\end{align}
Here, we defined a modified ponderomotive potential 
\begin{equation}
U_p'=\left(1+\frac{k_x}{c}\right) U_p
\label{Up_prime}
\end{equation}
with $U_p = A_0^2/4$ the ponderomotive potential in the dipole approximation. We note that the energy in equation \eqref{Up_prime} depends on the direction of asymptotic wavenumber of the electron through its projection on the laser propagation direction, $k_x$. The implications of such a  dependence was discussed in the context of a relativistic treatment in reference~\cite{Reiss1990}. The expression for $U_p'$ in equation \eqref{Up_prime} agrees to first order in $1/c$ with the result in equation (21) of reference \cite{Fritzsche2019}.  In equation \eqref{NDVGV} the quiver radius is defined as $\alpha_0= -A_0/\omega$ and the symbol $J_n(u,v)$ denotes a generalized Bessel function of integer order~\cite{Reiss1980}. The introduction of $J_n(u,v)$, through their generating function, facilitates the convenient form of the time-dependent phase in equation \eqref{NDVGV} at the cost of the infinite sum.

\subsection{PMD in the long-pulse limit}
In the $S$-matrix formulation~\cite{Reiss1980}, the leading-order transition for multiphoton ionization reads in the nondipole case
\be
\label{eq:S_multiphoton}
(S-1)^\text{B}_{fi} = -i \int_{-\infty}^\infty dt \langle\psi_\text{ND,VG}^{V,\bm k}
\lvert 
\bm A' \cdot \bm p + \frac{\bm A{'^2}}{2} 
\lvert
\phi_0\rangle.
\ee
where $|\psi_\text{ND,VG}^{V,\bm k}\rangle$ is given by equation \eqref{NDVGV}, $\bm A'$ is defined in equation \eqref{Aprime} and $|\phi_0\rangle$ is the field-free initial state, which includes the phase $e^{-i E_b t}$ describing the time evolution of the ground state with energy  $E_b=-I_p$. The superscript B indicates the leading-order Born expression for the $S$-matrix. We assume that the  pulse is switched on and off adiabatically at very early and late times and performing the time integral in equation \eqref{eq:S_multiphoton} results in
\be
(S-1)^\text{B}_{fi} = -2\pi  i  \sum_{n} T_{fi}(\bm k_n) \delta(\frac{k_{n}^2}{2} -n\omega + U_p' + I_p),
\label{MPI}
\ee
with an energy conserving delta function giving the final kinetic energy as 
\be
\frac{k_{n}^2}{2} =n\omega - U_p' -I_p
\label{energy_cons}
\ee
for $n > n_0$ and $n_0$ the smallest number that makes the right-hand side of equation \eqref{energy_cons} positive. The $T$-matrix element in equation \eqref{MPI} to first order in $1/c$ reads
\be
T_{fi}(\bm k_n) = (-i)^n (U_p' - n \omega) J_{-n}\left(\bm \alpha_0 \cdot \bm k_n, \frac{U_p'}{2\omega}\right) \tilde{\phi_0}(\bm k_n),
\ee
with $\tilde{\phi_0}(\bm k_n) = (2 \pi)^{-3/2} \int d \bm r \exp(-i \bm k_n \cdot \bm r) \phi_0(\bm r)$ the Fourier transform of the initial orbital evaluated at the momentum $\bm k_n$.
The multiphoton ionization rate is related to the norm square of the transition matrix~\cite{Reiss1980}. The energy $k_n^2/2$ for a given $n$ depends on the component of $\bm k$ along the propagation direction $\hat {\bm x}$ due to the presence of $U_p'$  in equation \eqref{energy_cons}.  We model the nondipole PMD by the norm square of the $T$-matrix element, i.e.,
\begin{equation}
\frac{dP}{d \bm k_n} = |(U_p' - n \omega)J_{-n}\left(\bm \alpha_0 \cdot \bm k_n, \frac{U_p'}{2\omega}\right) \tilde{\phi_0}(\bm k_n)|^2.
\label{PMD}
\end{equation}
In the case of the dipole approximation, the corresponding quantity reads
\begin{equation}
\frac{dP_\text{D}}{d \bm k_{n, \text{D}}} = |(U_p - n \omega)J_{-n}\left(\bm \alpha_0 \cdot \bm k_{n,\text{D}}, \frac{U_p}{2\omega}\right) \tilde{\phi_0}(\bm k_{n,\text{D}})|^2,
\label{PMDD}
\end{equation}
where the subscript D denotes that equation \eqref{PMDD} is only valid in the dipole approximation. We see from equations \eqref{PMD}-\eqref{PMDD} that the main differences between PMDs in the nondipole  and the dipole approximations  are the presence of $U_p'$ instead of $U_p$, and $\bm k_n$ instead of $\bm k_{n,\text{D}}$. We discuss the differences associated with the length of the momenta in section \ref{constraints}.

\subsection{Constraints on PMD from energy conservation}
\label{constraints}
In the dipole approximation,   the magnitude of the final momenta for $n$-photon absorption satisfy
\begin{align}
\frac{\knd^2}{2} = n\omega-U_p-I_p.
\label{eq:knd}
\end{align}
Equation \eqref{eq:knd} shows  that the possible final momenta  form a circle with its center in the origin and radius given by $\knd = \sqrt{2(n\omega-U_p-I_p)}$.  
If we consider the  $(k_x,k_z)$ plane ($k_y=0$), where the $x$-direction is the propagation direction and $z$ the polarization direction of the laser, the final momenta  read
\begin{align}
k_{n,\text{D},x}(\theta) &= \knd\cos(\theta),
\label{eq:kndx}\\
k_{n,\text{D},z}(\theta) &= \knd\sin(\theta),
\label{eq:kndz}
\end{align} 
where $\theta$ is measured from the $x$-axis.

Equation \eqref{energy_cons} constraints the final momenta in the  nondipole strong-field-approximation approach for the $n$'th photon absorption channel. We rewrite this expression, specializing to the  $(k_x,k_z)$ plane,  as
\begin{align}
(k_{n,x}+U_p/c)^2+k_{n,z}^2&= \knd^2+(U_p/c)^2.
\label{eq:knnd2}
\end{align}
Equation \eqref{eq:knnd2} shows that the final momenta in the nondipole long-pulse limit are confined to a circle  with center coordinates
\begin{equation}
(k_{x,0}, k_{z,0}) = (-U_p/c, 0),
\label{center}
\end{equation}
and with a radius 
\begin{equation}
k_n = \sqrt{\knd^2+(U_p/c)^2}.
\label{radius}
\end{equation}
Accordingly, in terms of the overall character of the energy-conserving ATI rings, the nondipole effect shifts the origin of the distribution by an amount $U_p/c$  opposite the propagation direction.  

If one wants to parameterize the nondipole momenta in the polarization and propagation plane in terms of  the angle measured from the origin and with respect to the  propagation direction, it can be done by assuming the form
\begin{align}
k_{n,x}(\theta) = k_{n}(\theta)\cos(\theta),
\label{eq:kxnnd}\\
k_{n,z}(\theta) = k_{n}(\theta)\sin(\theta)
\label{eq:kznnd},
\end{align}
where $k_{n}(\theta)$ is found by assuming the above form of the momenta components and inserting it into equation \eqref{eq:knnd2}. The angle-dependent length of the momentum is given as $k_{n}(\theta) = \sqrt{\knd^2 + U_p^2 \cos(\theta)^2/c^2}-U_p\cos(\theta)/c $, which to leading order in $1/c$ gives 
\begin{align}
k_{n}(\theta) = \knd-U_p\cos(\theta)/c. 
\label{eq:knnd_theta}
\end{align}
with the dipole momentum radius $\knd$ given below equation \eqref{eq:knd}. We see that the shifts in the length of the momenta, as described by equation \eqref{eq:knnd_theta}, are consistent with the shift of the origin of the nondipole momentum distributions. 

The relative importance of the nondipole shift decreases with the order of the photon-absorption or ATI channel, and hence the energy of the photoelectron. Namely, to leading order in $1/c$ and for large $n$, we obtain using equations \eqref{radius} and \eqref{eq:knd} that the ratio between the magnitude of the shift, $U_p/c$, and the radius, $k_n$, scales as $(U_p/c)/k_n \propto 1/\sqrt{n}$.

\section{Results and discussion}
\label{results}

To illustrate the nondipole effects, we consider a laser with a  wavelength of 3400 nm and an intensity of $\SI{6e13}{\watt\per\centi\meter^2}$. These values are similar to the those used in a recent experiment~\cite{Keller2014}, where nondipole shifts along the laser propagation direction were reported and are in a regime where the nondipole strong-field-approximation Hamiltonian is accurate~\cite{JensenPRA2020}. In the experiment a pulsed laser was used and the different photon absorption channels were not resolved. We focus on the long-pulse limit where the individual photon absorption channels can be resolved. The laser is linearly polarized in the $z$-direction and propagates in the $x$-direction. Ground state atomic hydrogen  is used as target. All results are obtained from equations \eqref{PMD} and \eqref{PMDD} and the energy relations of section II.

\begin{figure}
\centering
\includegraphics[scale=0.5]{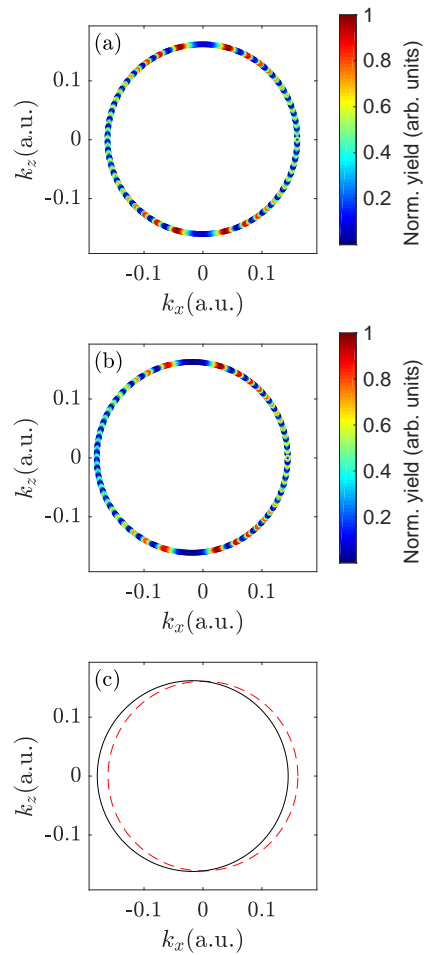}
\caption{(a) PMD for the $n=n_0=216$ absorption channel in the dipole approximation, corresponding to $n\omega=1.216U_p$. (b) As (a) but for the nondipole strong-field-approximation description. (c) Possible final momenta for the  nondipole (black, full) and dipole (red, dashed)  cases. In (a) and (b), the yields are normalized to the signal maxima.}
\end{figure} 

\begin{figure}
\centering
\includegraphics[width=0.45\textwidth]{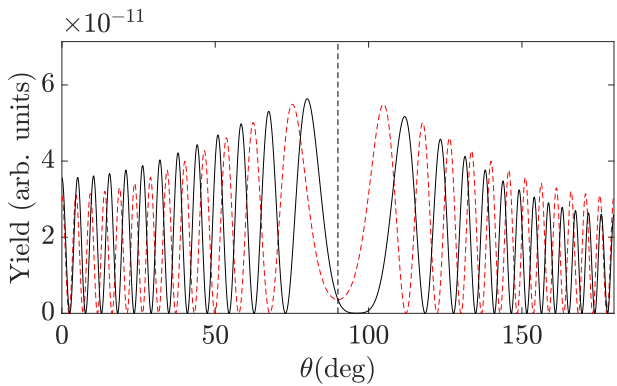}
\caption{Angular distribution for the absorption channel $n = n_0=216$ corresponding to $n\omega=1.216U_p$ for the nondipole (black, full curve) and dipole (red, dashed curve) cases. The angle $\theta$ is measured from the $k_x$-axis with respect to the origin. The vertical dashed line marks the polarization direction $\theta= 90^\circ$.}
\end{figure} 

\begin{figure}
\centering
\includegraphics[width=0.45\textwidth]{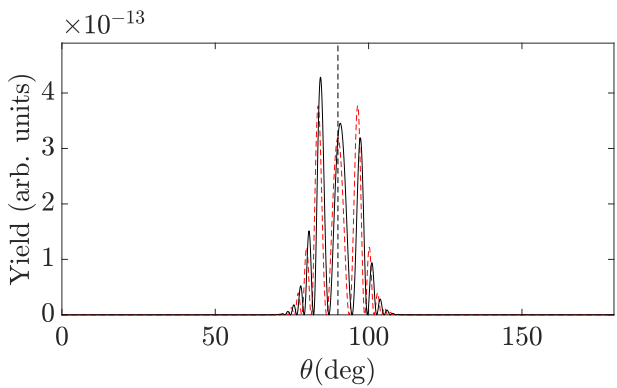}
\caption{As figure 2, but for absorption channel $n=n_0+83=299$, corresponding to $n\omega = 1.683U_p$.}
\end{figure}

Figure 1 shows the PMD in the plane spanned by the propagation and polarization directions of the laser for the first ATI peak, i.e., for the photon absorption channel with $n=n_0$, see equation \eqref{energy_cons}. Figure 1(a) shows the distribution in the dipole approximation, equation \eqref{PMDD}, while figure 1(b) shows the distribution obtained with the nondipole approach, equation \eqref{PMD}. Finally, figure 1(c) shows the energetically allowed final momenta with the dipole [equations \eqref{eq:kndx}-\eqref{eq:kndz}] and nondipole [equations \eqref{eq:kxnnd}-\eqref{eq:kznnd}] approaches. A comparison between the three panels illustrates the main characteristics of the nondipole effects in the long-pulse limit: a shift of the center of the energy-conserving rings  as given by equation \eqref{center}, i.e., a shift of the center along the negative $k_x$ axis by $U_p/c$ in the nondipole case compared to the dipole case.  The other main feature is a nondipole-induced redistribution of the continuum population along the allowed energy-conserving momentum values. This latter feature is highlighted in figure 2, which shows the angular distributions obtained from  the momenta in figure 1 for each direction $\theta$ with respect to the propagation direction, $x$.   The figure gives the results of the nondipole (full, black curve) and dipole (dashed, red curve)  approaches. We see that the angular distribution obtained within the dipole approximation is symmetric about the line of the laser polarization, dashed line at $\theta= 90^\circ$. The nondipole terms break this symmetry. We notice, for example, the nondipole-induced changes to the pair of dipole peaks closest to the polarization direction, i.e., closest to $\theta =90^\circ$. The nondipole terms induce an increase in the signal in the forward propagation direction ($\theta < 90^\circ$) and a decrease in the signal in the direction opposite to the propagation direction ($\theta > 90^\circ$). In a simple picture, we may think about this change as an effect of the radiation pressure. The cycle-averaged momentum kick in the $x$ direction is obtained from equation (20) of reference \cite{JensenPRA2020} and is given by  $U_p/c$ (see also references \cite{Reiss2013,Reiss2014} and references therein). A comparison of results like the ones in figures 1 and 2 for increasing values of $n$ (not shown) confirms that the nondipole-induced shifts decrease in relative magnitude for increasing ATI channel, i.e., as the photon absorption channel number $n$ increases. As an example, we show in figure 3 the result for $n=n_0+83=299$ , still within the range of the direct ATI electrons with energy less than $< 2 U_p$, but with much smaller nondipole-induced shifts in the peak positions although with some nondipole effects affecting the relative heights of the different peaks.
\begin{figure}
\centering
\includegraphics[width=0.45\textwidth]{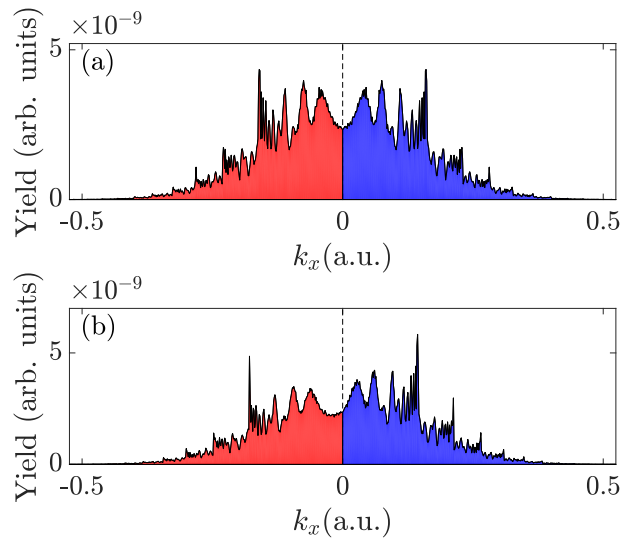}
\caption{Projection of the PMD from figure 1 onto the  $k_x$ axis for the (a) dipole  and (b) nondipole cases. The vertical dashed line marks the polarization direction $\theta= 90^\circ$.  The distribution is symmetric in $k_x$ in the dipole approximation, but asymmetric in the nondipole case. In (b) the area with $k_x < 0$ makes up 46.68$\%$ while the area with $k_x > 0$ makes up 53.32$\%$.}
\end{figure}

Finally, we illustrate that the ATI-channel-resolved nondipole-induced changes of the photoelectron distributions are also  present in spectra that are integrated over ATI channels.  Here, as an alternative to the shifting-of-the-peaks effect of figures 2 and 3, which is also present in the total angular distribution, we show the distribution summed over $n$-channels as a function of the momentum $k_x$ in the laser propagation direction. We see from figure 4(a) , that the distribution, as expected, is symmetric around $k_x=0$ in the case of the electric dipole approximation. In the nondipole approach, an asymmetric distribution is predicted as shown in figure 4(b). We observe that the present nondipole approach predicts that most of the strong-field ejected electrons, ionize into states with positive momenta along the laser propagation direction. 

\section{Conclusion}
In the present work, we considered nondipole effects in the PMD following  atomic strong-field ionization by an intense laser field with a wavelength in the mid-infrared regime. We focused on the long-pulse limit, where the individual ATI channels can be resolved.  In this limit we showed that the nondipole effect leads to two prominant effects. First, in the plane formed by the propagation and polarization directions, the nondipole effects lead to a shift of the center of the energy-conserving ring for a given ATI channel in the direction opposite to the laser propagation direction with a magnitude of $U_p/c$. Second, the nondipole effects lead to a redistribution of the continuum electron population on a given ring. We confirmed a nondipole-induced shift in the distribution of the momenta such that the PMD is no longer symmetric about the axis of laser polarization. The nondipole terms induce a shift in the population towards momenta with a positive component in the propagation direction. This shift is consistent with a radiation-pressure effect, but its direction can be affected by Coulomb effects~~\cite{Madsen2006,Bauer2017,Haram2019,Fritzsche2019,Doerner2019}, not considered here. The relative importance of the nondipole-induced shift and continuum population redistribution decreases as a function of the order $n$ of the ATI channel and hence decrease as a function of the energy of the photoelectron. Since the signal from the relative low orders of $n$ dominate over that from the higher order, there is still a significant nondipole effect in signals that are summed-up over the contributions from different ATI channels.  The considered long-pulse limit provides a supplementary energy-resolved perspective on nondipole effects in strong-field ionization of matter.

\begin{acknowledgments}
This work was supported by the Danish Council for Independent Research (Grant No. 9040-00001B).
\end{acknowledgments}

%\bibliography{referencer}

\end{document}